\documentstyle[12pt]{article}
\title{ \bf Modelling Meso-Scale Diffusion Processes in Stochastic Fluid
Bio-membranes}
\author{H. Rafii-Tabar$^1$\thanks{Corresponding author. H. Rafii-Tabar, School of
Computing and Mathematical Sciences, University of Greenwich, Woolwich
Campus, Wellington Street, London SE18 6PF, UK. Tel: (+44)0181-3318548. Fax:
(+44)0181-3318665. Email: h.rafii-tabar@gre.ac.uk}\,\, and H.R. Sepangi$^2$
\\  \small $^{1}$ Computational Nano-Science Research Group, Centre for
Numerical Modelling and Process Analysis,\\  \small School of Computing and
Mathematical Sciences, University of Greenwich,\\  \small Woolwich Campus,
Wellington Street, London SE18 6PF, UK. \\ \small
$^{2}$  Department of Physics, Shahid Beheshti University, Evin,
Tehran 19834,  Iran.} 
\begin{document}
\maketitle
\vspace{11cm}
\pagebreak
\begin{abstract}
The space-time dynamics of rigid inhomogeneities (inclusions) free to move in a
randomly fluctuating fluid bio-membrane is derived and numerically simulated as a
function of the membrane
shape changes. Both vertically placed (embedded) inclusions and horizontally
placed (surface) inclusions are considered. The energetics of the membrane, as a
two-dimensional (2D) meso-scale continuum sheet, is described by the
Canham-Helfrich
Hamiltonian, with the membrane height function treated as a stochastic process.
The diffusion
parameter of this process acts as the link coupling the membrane shape
fluctuations to
the kinematics of the inclusions. The latter is described via Ito stochastic
differential
equation. In addition to stochastic forces, the inclusions also experience
membrane-induced deterministic forces. Our aim is to simulate the diffusion-driven
aggregation of inclusions and show how the external inclusions arrive at the sites
of
the embedded inclusions. The model has potential use in such emerging fields as
designing a targeted drug delivery system. \\ 
\vspace{5mm}
\end{abstract}
\pagebreak
Amphiphilic molecules, such as lipids and proteins, can 
self-assemble themselves into a variety of exotic structures in an aqueous
environment \cite{Hunter}. Computer-based simulation of the dynamics of these
structures
forms an interesting research area in the computational statistical mechanics of bio-
material systems.  One such structure is the phospholipid bilayer which represents
the generic structure of all bio-membrane systems, both natural and artificial.
These membranes can have thicknesses of only few nano-metres but linear sizes
of up to tens of micro-metres, and can therefore be regarded as highly flexible,
fluid-like, 2D continuum sheets embedded in a three-dimensional space. 

Thermal fluctuations can induce shape fluctuations and shape transformations in
membranes. For example, in the so-called budding transition \cite{Seifert} a
spherical vesicle transforms
into a prolate ellipsoid as the temperature increases. There is also the possibility
that the spherical geometry becomes oblate, producing a shape similar to the
biconcave rest shape of a red blood cell. 

Bio-membranes regulate the recognition and transport processes to and from the
interior of the cells, as well as between the cells, forming a barrier which all
external particles arriving at the cell must cross. They contain a variety of
integral (embedded) inhomogeneities, such as proteins and other macromolecules
\cite {Singer}, that penetrate the thickness of the membrane and act as transport
channels. We shall refer to these as the M-type inclusions. These inclusions are
mobile and can freely diffuse across the membrane. Their presence, however, force
the bilayer to adjust its thickness locally so as to match the
thickness of the hydrophobic region of these inclusions \cite{Dan1,Dan2}, causing
local
deformations in the membrane geometry.  The perturbations produced in
the membrane shape due to these local deformations give rise to both
short-range and long-range {\it membrane-induced} indirect forces between the
inclusions. These forces act in addition to direct Van der Waals and electrostatic
forces between the inclusions. The Long-range forces originate from the
perturbations associated with the long wavelength shape fluctuations
\cite{Goulian}, whereas the short-range forces are associated with the local
deformations in the immediate vicinity of the inclusions \cite {Dan1}. These
membrane-induced forces between the inclusions play a far more significant role,
viz the direct molecular interactions, when the length scales involved are
comparable to the size of the membrane.  In addition to this mode of deformation,
a membrane can also deform due to the tension at the amphiphilic molecules-water
interfaces. This tension results in a change in the overall surface area of the
membrane. A third mode of deformation also exists and this is associated with the
{\it bending} elastic-curvature property of the membrane which distinguishes it
from a sheet of simple fluid dominated by surface tension \cite{Seifert}.
Accordingly, two models to study the inclusion-induced local deformations have
been developed. In the first model, the membrane energy is taken to consist of a
contribution from the molecular expansion/compression due to the change in the
thickness at the inclusion boundary, and also a contribution from the overall change
in the surface area. Using this
model, it is shown that \cite {Bloom,Abney,Marcelja,Owicki,Fattal} the inclusion-
induced deformations cause {\it exponential} decays in the thickness of the
membrane, extending from the inclusion-imposed value to the
equilibrium thickness value, as shown schematically in Fig.1 for two rod-like
inclusions. In the second model
\cite{Dan1,Huang}, the contribution of the membrane bending property is taken
into account in the energy term, and it is found that this significantly affects the
perturbation profile at the inclusion boundary as well as modifying the membrane-
induced interactions. Evidently, an object supported by surface tension would have
a different dynamics than one supported by the bending elasticity of the surface
\cite{Mansfield}.  

In addition to the M-type inclusions, membranes can also carry inclusions that lie
on their surfaces \cite{Golestanian} as shown schematically in Fig.2. These surface
inclusions can represent objects that have arrived at the membrane from the outside
and are therefore referred to as external inclusions. We shall refer to these as the
S-type inclusions.

At the meso-scale, i.e. when the detailed molecular architecture of the membrane
can be subsumed into a background 2D sheet, the free elastic energy of a
symmetric membrane is described by the Canham-Helfrich Hamiltonian
\cite{Canham,Helfrich}
\begin{equation}
{\cal H}=\int d^2\sigma\sqrt{g}\{\sigma_0+2\kappa H^2+\bar{\kappa}K\}\,,  
\label{eqn00}
\end{equation}
where
\begin{eqnarray}
K=\mbox{det}(K_{ij})=\frac{1}{R_1 R_2}\,,\nonumber\\  
H=\frac{1}{2}\mbox{Tr}(K_{ij})=\frac{1}{2}\left(\frac{1}{R_1}+\frac{1}{R_2}
\right). 
\label{eqn01}
\end{eqnarray}
are respectively the Gaussian and mean curvatures of the sheet, $R_1$ and $R_2$
are the two principle radii of curvature of the sheet, $\sigma_0$ is the surface
tension, $\kappa$ is the bending rigidity, $\bar{\kappa}$ is the Gaussian rigidity,
$g$ is determinant of the metric tensor and $\sigma=(\sigma_1,\sigma_2)$ is the
2D local coordinate on the sheet as opposed to the coordinates on the embedding
space. The last term in (\ref{eqn00}) is, by
Gauss-Bonett theorem, an invariant for closed surfaces implying that the dynamics
of a membrane is not influenced by this term if its topology  remains fixed. In what
follows, we concentrate on
membranes with {\it fixed topology} and drop this term. We then have
\begin{equation}
{\cal H}=\int d^2\sigma\sqrt{g}\{\sigma_0+2\kappa H^2\}. 
\label{eqn02}
\end{equation}
The study of a membrane whose free energy is described by (\ref{eqn02}) is
facilitated by considering it to be nearly flat, i.e. its thickness to be much smaller
than its linear size $L$. This is indeed what we mean by a meso-scale model of a
membrane.
We therefore take the membrane to be almost parallel to the $(x_1,x_2)$ plane,
regarded as the reference plane. The position of a point on
the membrane  can then be described by a single-valued function $h(x_1,x_2)$
representing the {\it height} of that point. This simplification is achieved by writing
the Hamiltonian (\ref{eqn02}) in the Monge representation \cite{Gompper} which 
gives for the mean curvature
\begin{equation}
2H=-g^{-3/2}[\partial_1^2 h(1+(\partial_2 h)^2)+\partial_2^2 h(1+(\partial_1
h)^2)-2\partial_1 h\partial_2 h\partial_1\partial_2 h]\,,  \label{eqn03}
\end{equation}                                                     
where $\partial_i\equiv\frac{\partial}{\partial x^i},\hspace{2mm} i=1,2$.
We assume that the area of the membrane can fluctuate without constraint by
setting $\sigma_0=0$ in (\ref{eqn02}). Consequently, using (\ref{eqn03}), the
Hamiltonian (\ref{eqn02}) to leading order in derivatives of $h$ becomes 
\begin{equation}
{\cal H}_0=\frac{\kappa}{2}\int d^2 x(\nabla^2 h)^2.  \label{eqn1}
\end{equation} 
This is the Canham-Helfrich Hamiltonian in Monge representation, expressed in
terms of the height function of the membrane. It is the expression that we employ
to describe the energetics of the membrane.

Employing a statistical mechanics based on (\ref{eqn1}) only, and ignoring the
contributions from the expansion/compression and interfacial energies, the potential
energy function
\begin{equation}
V_{MM}^{T}(R_{ij})=-k_B T \frac{12 A^2}{\pi^2 R_{ij}^4}\,, \label{eqn2}
\end{equation}
was constructed \cite{Goulian} to describe the membrane-induced temperature-
dependent long-range forces between a pair of disk shape M-type inclusions that
can freely tilt with respect to each other. Another function was also constructed
for long-range interaction between two S-type inclusions \cite{Golestanian}
\begin{equation} 
V_{SS}^{T}(R_{ij},\theta_i,\theta_j)=-k_B T \frac {L_i^2 L_j^2}{128 R_{ij}^4}
\cos^2[2(\theta_i + \theta_j)]\,, \label{eqn3}
\end{equation}
where $A=\pi r_{0}^2$ is the area of an M-type inclusion of radius $r_0$, $k_B$
is the
Boltzmann constant, $R_{ij}$ is distance between the centres of mass of two
inclusions $i$ and $j$, $L_i$ and $L_j$
are the lengths of two S-type inclusions making the angles $\theta_i$ and
$\theta_j$
respectively with the line joining their centres of mass (see Fig.2) and $T$ is the
membrane temperature. It is evident that both of these membrane-induced
potentials are attractive and fall off
as $R^{-4}$ with the distance. These expressions are derived for rod-like inclusions
that are assumed to be much more rigid than the ambient membrane so that these
inclusions can not move coherently with the membrane. The only  degrees of
freedom for the rods are rigid translations and rotations while they remain attached
to the membrane. 
 
So far, the modelling of bio-membrane dynamics decorated with inclusions has
been mainly concerned with constructing potential energy functions such as those
given in (\ref{eqn2}) and (\ref{eqn3}). An interesting problem, however, would be
to use this information to simulate the space-time behaviour of inclusions in a
membrane described
by (\ref{eqn1}) and undergoing stochastic shape fluctuations. This is the problem
that we address in this paper. This type of simulation can establish a direct link 
between the randomly changing membrane shape on the one hand
and the inclusion dynamics on the other. In such a simulation, the thermodynamic
phase behaviour of inclusions, such as their temperature-dependent aggregation,
can be directly computed. This phase behaviour plays a crucial role in the
functional specialisation of a membrane \cite{Dan1}. Furthermore, information on
the capture rate of the S-type inclusions,
which could represent external drug particles, at the sites of the M-type inclusions
can be obtained as a function of the changes in the environmental variables such
as the ambient temperature. This type of meso-scale simulation when coupled with
the
Molecular Dynamics (MD) simulation of membrane patches near the inclusions at
the nano-scale \cite{Berger}, can produce a seamless {\it multi-scale}
model of the entire environment for many bio-molecular processes, starting with
the arrival of external inclusions at the cell, their diffusion in the membrane, and
finally their molecular docking at the site of the embedded inclusions.

To proceed, let us consider a 2D bio-membrane described by (\ref{eqn1})
containing both the
M-type and the S-type inclusions. To make the membrane a stochastically
fluctuating
medium, we treat the height function in (\ref{eqn1}) as a stochastic Wiener process
with a Gaussian distribution, whose mean and variance can be written
as \cite{Gardiner}
\begin{equation}
\langle h(x_1,x_2;t) \rangle=0\,, \label{eqn4}
\end{equation}
\begin{equation}
\langle h(x_1,x_2;t)\,h(x_1,x_2;t) \rangle =\langle [h(x_1,x_2;t)-\langle h(x_1,x_2;t)
\rangle]^2 \rangle\,=2Dt\label{eqn5}
\end{equation}
where $D$ is the diffusion constant associated with the height fluctuations at
the local position $(x_1,x_2)$ and represents the measure with which random
fluctuations
propagates in the local geometry. Such random height changes would cause a
roughening of the membrane surface on molecular scales, and this has been
observed in NMR experiments \cite {Chiu}. 

Assuming that this is the only stochastic process present in the membrane, it is
then reasonable to suppose that this stochastic dynamics is communicated to the
inclusions as well, and that their ensuing random motions are contingent only on
these fluctuations. This implies that the mathematical point representing the centre
of mass of an inclusion coinciding with the membrane point $(x_1,x_2)$ would also
experience the same fluctuations and would diffuse with the same diffusion
constant. To derive an
expression for $D$, based on (\ref{eqn1}), we start with the {\it static}
height-height correlation function
obtained from (\ref{eqn1}). This is given by \cite{Seifert,Safran} 
\begin{equation}
\langle h({\bf q};0) h^\ast({\bf q^\prime};0)\rangle=\frac{k_B T}{\kappa
q^4}(2\pi)^2\delta({\bf q}-
{\bf q}^\prime)\,,\label{eqn6}
\end{equation}
where $\langle\cdots\rangle$ is the thermal averaging with respect to the
Boltzmann weight, 
$\mbox{exp}(-{\cal H}_0/k_B T)$, and $\bf q$ is the wave vector of magnitude
$q$.
The corresponding {\it dynamic} correlation function can be obtained \cite{Seifert}
by writing 
\begin{equation}
h({\bf q};t)=h({\bf q};0)e^{-\gamma_0({\bf q})t}\,,  \label{eqn7}
\end{equation}
giving
\begin{equation}
\langle h({\bf q};t) h^\ast({\bf q^\prime};t)\rangle=\frac{k_B T}{\kappa q^4}
e^{-2\gamma_0(q) t}(2\pi)^2 \delta({\bf q}-{\bf q}^\prime)\,,\label{eqn8}
\end{equation}
where the damping rate, $\gamma_0(q)$, reflecting the long-range character of the 
hydrodynamic damping, is defined as
\begin{equation}
\gamma_0(q)=\kappa q^3/4\eta\,, \label{eqn9}
\end{equation}
and $\eta$ denotes the coefficient of viscosity of the fluid membrane. In real
space, (\ref{eqn8}) transforms \cite{Gompper,Safran} to 
\begin{equation}
\langle h(x_1,x_2;t)h(x_1,x_2;t)\rangle = \frac{k_B T}{4\pi\kappa}L^2
e^{-2\gamma_0t}\,,\label{eqn10}
\end{equation}
where $L$ is the length of the membrane. This is the equal-time correlation
function for membrane fluctuations. A similar model of an {\it active} fluctuating
membrane in which the vertical displacements of the membrane satisfy a Langevin
equation in the $\bf q$ space has also been proposed \cite {Prost}, and is its
shown that a term similar to the static version of (\ref{eqn8}) contributes to the
correlation function which also contains a contribution from  non-equilibrium 
fluctuations. The latter is in the form of a $q^{-5}$ term which dominates at long
distances. Comparison of (\ref{eqn5}) and
(\ref{eqn10}) yields the desired result
\begin{equation}
D=\frac{\left(\frac{k_B T}{4\pi\kappa}\right)L^2 e^{-2\gamma_0 t}}{2
t}.\label{eqn11}
\end{equation}
It should be emphasised that the association of a diffusive process with the
membrane height function, and the resulting diffusion constant, is not analogous
to the usual model of a diffusion process in which, for example, a particle diffuses
through a medium, such as a fluid. Rather, what is suggested here is that the {\it
magnitude} of a mathematical function representing the height of a mathematical
point in the membrane is subject to random stochastic variations, and the diffusion
constant is a measure of this variation.    

When the M-type inclusions are present they produce exponentially decaying local
deformations in the membrane geometry (see Fig.1). Correspondingly, the
correlation function can be modified by a multiplicative exponential factor to 
\begin{equation}
\langle h(x_1,x_2;t) h(x_1,x_2;t)\rangle_{NI}= e^{-r_0/R}\langle h(x_1,x_2;t)
h(x_1,x_2;t)\rangle \,,\label{eqn12}
\end{equation}
where $r_0$ is the radius of an M-type inclusion, $R+r_0$ is the radius of the
circular region around the inclusion with its centre coinciding with that of the
inclusion, and $NI$ stands for near inclusion. It is evident that outside this region
the exponential decay of the profile is negligible. Accordingly, within this circular
region of radius $R+r_0$, the diffusion constant is also modified to
\begin{equation}
D_m=De^{-r_0/R}. \label{eqn13} 
\end{equation}
This equation implies that when the centre of mass an S-type inclusion enters a
circular region of radius $R+r_0$ its diffusion coefficient goes over to $D_m$ and
progressively approaches zero as the boundary of an M-type inclusion is
approached. We can ascertain, as a first approximation, that this is how an M-type
inclusion interacts with an S-type inclusion.

In our simulations, the equations motion of both the S-type and the M-type
inclusions are represented by the differential equation of the
the Ito stochastic calculus \cite{Gardiner}
\begin{equation} 
d{\bf r}(t)={\bf A}[{\bf r}(t),t]\, dt+D^{1/2}d{\bf W}(t). \label{eqn14}
\end{equation}
This equation describes the stochastic trajectory, ${\bf r}(t)$, of the centres of
mass of the inclusions in terms of a dynamical variable of the inclusions, ${\bf
A}[{\bf r}(t),t]$, which is referred to as the drift velocity, and a term, $d{\bf
W}(t)$, which  is a
given Gaussian Wiener process with the mean and variance given by
\begin{eqnarray}
\langle d{\bf W}(t) \rangle=0 \nonumber\\
\label{eqn15} \\
\langle d{\bf W}_i(t)d{\bf W}_j(t) \rangle=2 \delta_{ij}dt. \nonumber
\end{eqnarray}
Equation (\ref{eqn14}) applies to each dimension of the motion. The Ito equation
predicts the increment in position, i.e. $d{\bf r}(t)={\bf r}(t+dt)-{\bf r}(t)$, for a
meso-scale time interval $dt$ as a combination of a {\it deterministic} drift part,
represented by ${\bf A}[{\bf r}(t),t]$, and a {\it stochastic} diffusive part
represented by $
D^{1/2} d{\bf W}(t)$ and {\it superimposed} on this drift part. This equation
resembles the
`position' Langevin equation describing the Brownian motion of a
particle \cite{Allen}. The position Langevin equation corresponds to the {\it
long-time} (diffusive time) 
configurational dynamics of a stochastic particle in which its momentum
coordinates are in thermal equilibrium and hence have been removed from the
equations of motion. Since we are interested in diffusive time scales as well, we
can re-write
(\ref{eqn14}) as
\begin{equation}
d{\bf r}(t)= \frac{D}{k_B T} {\bf F}(t)\,dt+D^{1/2} d {\bf W}(t)\,, \label{eqn16}
\end{equation}
where ${\bf F}(t)$ is the instantaneous systematic force experienced by the $i$-th
inclusion and is obtained from the inter-inclusion potentials, given in (\ref{eqn2})
and (\ref{eqn3}), according to
\begin{equation}
{\bf F}_i=-\sum_{j>i}{\bf \nabla}_{{\bf R}_i} V(R_{ij}). \label{eqn17}
\end{equation}
We implemented (\ref{eqn16}) for our 2D simulations according to the iterative
scheme\cite{Rafii} 
\begin{eqnarray}
X(t+dt)=X(t)+\frac{D}{k_B T}F_{X}(t)\,dt+\sqrt{2D dt}\,\, R^{G}_{X}
\nonumber\\
\label{eqn18} \\
Y(t+dt)=Y(t)+\frac{D}{k_B T}F_{Y}(t)\,dt+\sqrt{2D dt}\,\, R^{G}_{Y} \nonumber
\end{eqnarray}
where $R^{G}_{X}$ and $R^{G}_{Y}$ are standard random Gaussian variables
chosen separately and independently for each inclusion according to the procedure
given in \cite{Allen}, and $F_X,F_Y$ are the $X$ and $Y$ components of the force
$\bf F$. For the S-type inclusions, we treated the angles in (\ref{eqn3}) as
independent stochastic
variables described by 
\begin{equation}
\theta(t+dt)=\theta(t)+ \frac {D}{k_B T L^2} \tau(t)\,dt +\frac {1}{L} \sqrt{2D
dt}\,\, \theta^{G}\,, \label{eqn19}
\end{equation}
where $\tau$ is the torque experienced by an S-type inclusion and is given by
\begin{equation}
\tau_i=-\sum_{j>i} \frac{\partial V^T(R_{ij},\theta_i,\theta_j)}{\partial \theta_i}\,,
\label{eqn20}
\end{equation}
and $\theta^{G}$ is the angular counterpart of $R^{G}_{X}$ and $R^{G}_{Y}$.

In the numerical simulations, recently reported in their broad outlines \cite {Rafii1},
we use a square membrane with $L=40 \mu$m on its side. The other parameters
used were set at $\kappa=10^{-19}$ J and $\eta=10^{-3}$ J sec m$^{-3}$
\cite{Seifert}. These values correspond to the condition in which
the bending mode of the membrane is important. From these data the damping
coefficient, $\gamma_0$, in the real space of the membrane, can be obtained from
(\ref{eqn9}). The simulation temperature was set at $T=300^\circ\,K$, and the
correlation
(delay) time, $t$, over which the diffusion coefficient in (\ref{eqn11}) was
calculated, was set at $t=10^{-4}$sec. These data gave $D=2.6\times\,10^{-9}
$ m$^2$sec$^{-1}$, which is in close agreement with the value of $D\approx
4.4\times\,10^{-9}$m$^2$sec$^{-1}$ obtained at the molecular level via an MD
simulation of a fully hydrated phospholipid
dipalmitoylphosphatidylcholine (DPPC) bilayer diffusing in the $z$-direction
\cite{Tieleman}. To justify our choice of the correlation time, $t=10^{-4}$ sec,
we recall that the time scale of a stochastic particle, $t_D$, of mass $m$ is usually
determined from the relation \cite{Dhont} 
\begin{eqnarray}
t_D=\frac{m\,D} {k_B\,T} \label{eqn21} 
\end{eqnarray}
Since $t_D$ is normally of the order of $10^{-9}$sec, then for the criterion of
long-time dynamics, employed in our model (cf (\ref{eqn16})), to be justified the
correlation (diffusive) time scale, $t$, in (\ref{eqn11}) has to satisfy the condition
\cite{Dhont}
\begin{eqnarray}
t\gg t_D. \label{eqn22} 
\end{eqnarray}
For our calculated value of $D$ and our choice of the inclusion mass $m=10^{-12}
\mu$g, corresponding to an inclusion of length $L_i= 0.1 \mu$m, we obtained a
value of $t_D=0.6 \times 10^{-9}$sec, showing that our choice of the correlation
time was appropriate to satisfy the condition in (\ref{eqn22}). The radius of
an M-type inclusion was set at $r_0=0.01{\mu}m$, and the inclusions were all
equal in length.
  
The stochastic trajectories of the inclusions were obtained in a set of five
simulations. The simulation time step, $dt$, in (\ref{eqn18}) was set at
$dt=10^{-9}$sec, and each simulation was performed for $4\times\,10^{6}$ time
steps, i.e for a mesoscopic interval of $4000 \mu$sec. The total number of
inclusions considered was $36$, consisting of $13$ S-type and $23$ M-type. 

In the first simulation, we computed the random motions of the S-type inclusions
in a
membrane devoid of the M-type inclusions. This was done in order to observe the
details of the drift-diffusion motion over mesoscopic time scales. Figure 3 shows
the stochastic
X-Y trajectories of a sample of 4 S-type inclusions plotted on a {\it
micron}
scale up to the end of the simulation time. In addition to the drift motions,
represented
by the second terms in (\ref{eqn18}), the random
Brownian-type variations, emanating from the membrane shape fluctuations, are
superimposed on this drift motion and are clearly visible over the mesoscopic
length and time scales. 
Figure 4 shows the snapshots of a small patch of the membrane with
both the S-type (white spheres) and the M-type (black spheres) inclusions. In this,
and subsequent figures, the solid spheres refer
to the centres of mass of the rod-like inclusions. In the initial state the outer M-type
inclusions were regularly positioned, whereas the inner ones were
randomly distributed. The S-type inclusions were all distributed completely at
random. Figures 4a to 4c refer to the
simulation in which the M-type inclusions were pinned to the membrane, i.e. were
static, and only the S-type inclusions were mobile, and figures 4d to 4f refer to the
simulation in which both the M-type and the S-type inclusions were mobile. The
initial states in both simulations, figures 4a and 4d,
were the same. The snapshots were obtained from dynamic simulations, akin to 
a MD simulation, covering the entire
simulation time interval. These snapshots were recorded at $2\times 10^{-3}$ sec
intervals, with  figures 4c and 4f referring to the final states reached at the
conclusion of the simulations after $4\times\,10^{6}$ time steps. The animation
of a complete run showed clearly the stochastic motions of the inclusions, and how
the S-type inclusions approached the M-type inclusions and were captured at the
site of the M-type inclusions. An examination of
figure 4 shows that for the case of dynamic M-type inclusions, a larger
number of the S-type inclusions were captured at the M-type inclusion sites, i.e.
the number was some 4 times higher than in the static case at the same
temperature. We adopted the method of counting an S-type inclusion as a captured
inclusion when its centre of mass coincided with that of an M-type inclusion. The
numerical algorithm then transformed the colour code of that S-type inclusion from
white to black. Figures 4e and 4d also show
some diffusion-driven local {\it aggregation} of the M-type inclusions. 
The capture of the S-type inclusions can be viewed as the first stage in the
molecular docking process which will eventually transfer these inclusions into the
interior of the cell.

To examine the membrane response to temperature changes, two of the
simulations were performed at different temperatures. Figure 5 shows the
snapshots of these simulations
at $T=100^\circ$K (a to c) and at $T=350^\circ$K (d to f). Figures 5d to 5f
clearly show that both the number of captured
inclusions and the aggregation of the M-type inclusions were affected by these
temperature differences, as can be seen by comparing figures 5c
and 5f.

To sum up, although many dynamical aspects of membrane-like surfaces have been
addressed in the past \cite{Nelson}, it is only relatively recently that attention has
focused on the dynamics of membranes with inclusions. To our knowledge no
computer-based simulation of this dynamics has been reported so far.  In this paper
we constructed a meso-scale model of a generic bio-membrane
based on the Canham-Helfrich curvature-energy formalism. We treated the
height function of the membrane as a stochastic Wiener process whose correlation
function provides the relevant diffusion constant describing the membrane
fluctuations. Two types of inclusions, one mimicking the internal embedded type
and the other the external floating type, are carried by this membrane. These
inclusions experience the same stochastic fluctuations as those experienced by the
membrane itself, resulting in the transformation of their deterministic (drift)
space-time dynamics into a stochastic Langevin-type dynamics described by the
Ito stochastic calculus.
A set of {\it dynamic} simulations, resembling the standard MD simulations, were
performed to investigate the phase behaviour of these inclusions. In addition
to stochastic forces, these inclusions also experience deterministic interactions 
described by inter-inclusion potentials varying as $1/R^4$ with their separations.
The simulation results clearly indicate that the capture and aggregation rates
change with the temperature and that the embedded {\it mobile} inclusions capture
a greater number of the floating inclusions. A further extension of the present work
would be to include the influence of the surface tension, as well as the bending
rigidity, by keeping the corresponding term in (\ref{eqn02}). This will constrain the
fluctuations in the surface area of the membrane and would have a direct bearing
on the inclusion dynamics.      
\vspace{5mm}\\
The second author(HRS) is grateful to the UK Royal Society for financial support
through a visiting research fellowship and to the School of Computing and
Mathematical Sciences (Greenwich University) for their hospitality. Both authors
acknowledge useful discussions with Professor E. Mansfield FRS on the dynamics
of objects supported by surface tension.

\pagebreak

{\large {\bf Figure captions}}\vspace{10mm}\\
Figure 1: Two rod-like embedded (M-type) inclusions vertically placed in an
amphiphilic fluid membrane. The inclusions impose exponentially decaying
thickness-matching constraints on the bilayer at the inclusion boundary. Heavy
solid lines represent amphiphilic molecules. Figure based on \cite{Dan2}.
\vspace{5mm}\\
Figure 2: Two rod-like surface (S-type) inclusions lying on the surface of the
membrane. The rods have lengths $L_1$ and $L_2$, widths $\epsilon_1$ and
$\epsilon_2$ and making angles $\theta_1$ and $\theta_2$ with the line joining
their centres of mass. Figure based on \cite{Golestanian}.\vspace{5mm}\\ 
Figure 3: A small patch of the membrane showing the stochastic X-Y trajectories
obtained from equation (\ref{eqn18}) for a sample of four S-type inclusions without
the presence of the M-type inclusions. Both the drift and diffusion motions can be
clearly distinguished. \vspace{5mm}\\
Figure 4: A set of snapshots, obtained from dynamic simulations, showing the
capture of rod-like S-type inclusions (white spheres) at the rod-like M-type inclusion
sites (black spheres) for static (a to c) and dynamic (d to f) M-type inclusions. The
aggregation of the M-type inclusions can also be observed (d to f). Only the centres
of mass of the inclusions are shown.\vspace{5mm}\\
Figure 5: A set of snapshots, obtained from dynamic simulations, showing the
capture of rod-like S-type inclusions at the rod-like M-type inclusion sites for mobile
M-type inclusions at $T=100^\circ$K (a to c) and $T=350^\circ$K (d to f). Only
the centres of mass of the inclusions are shown.
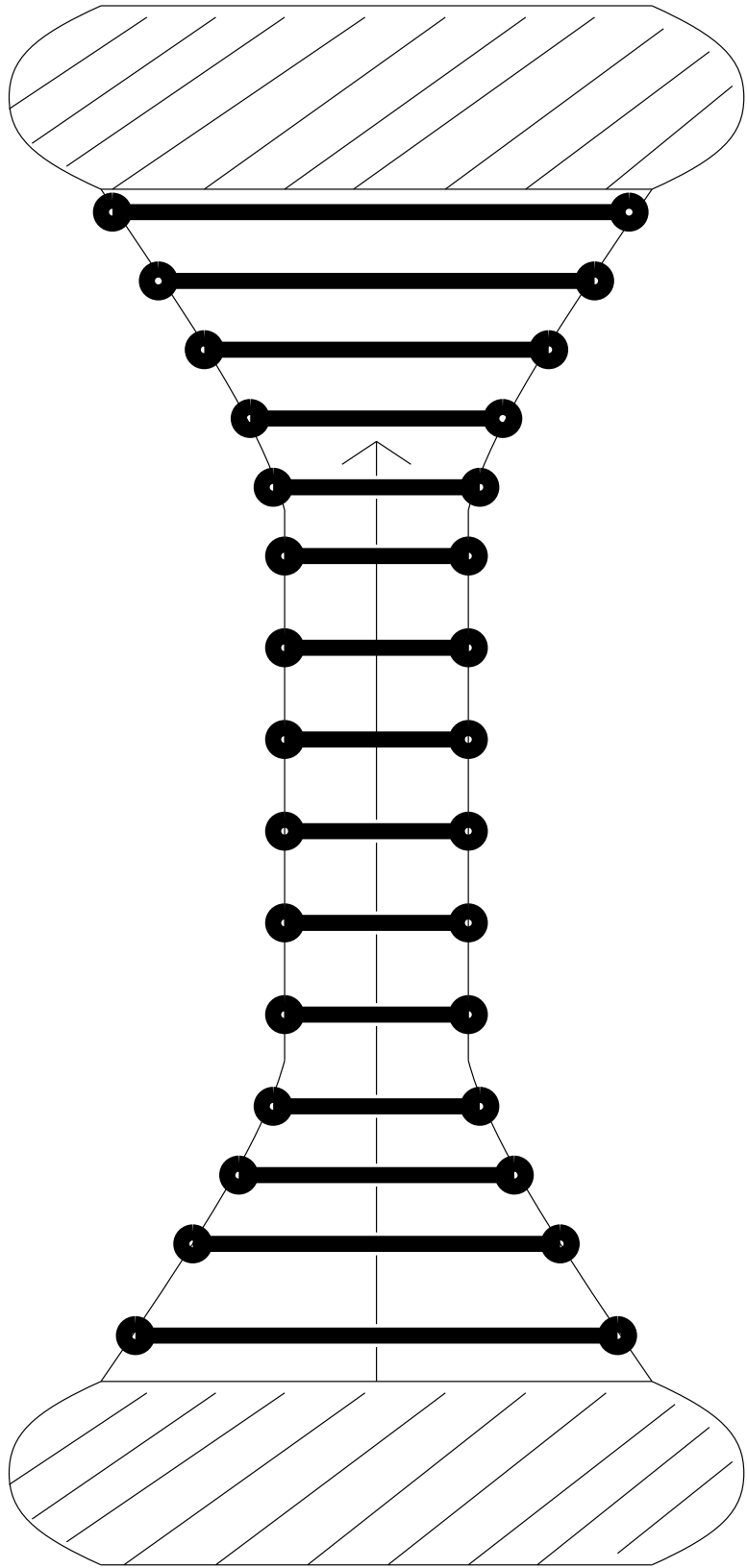
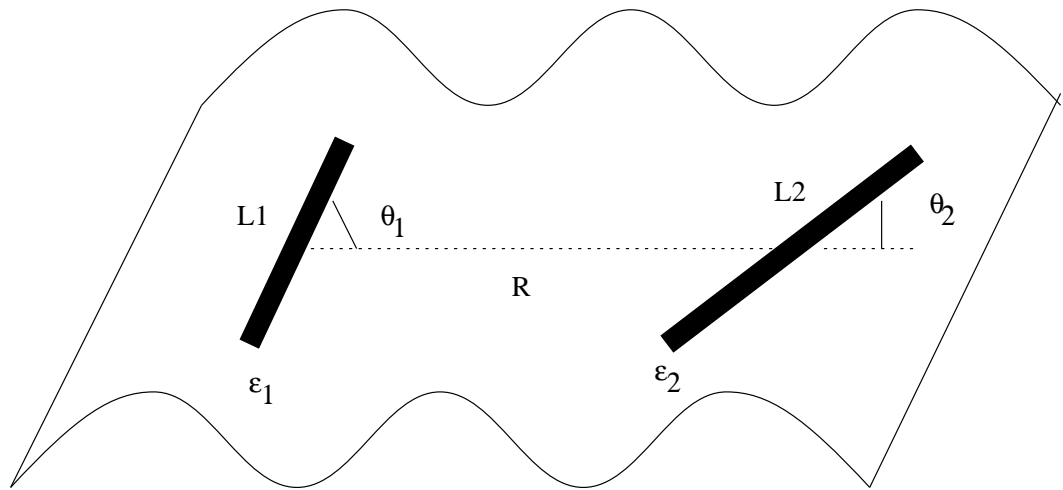
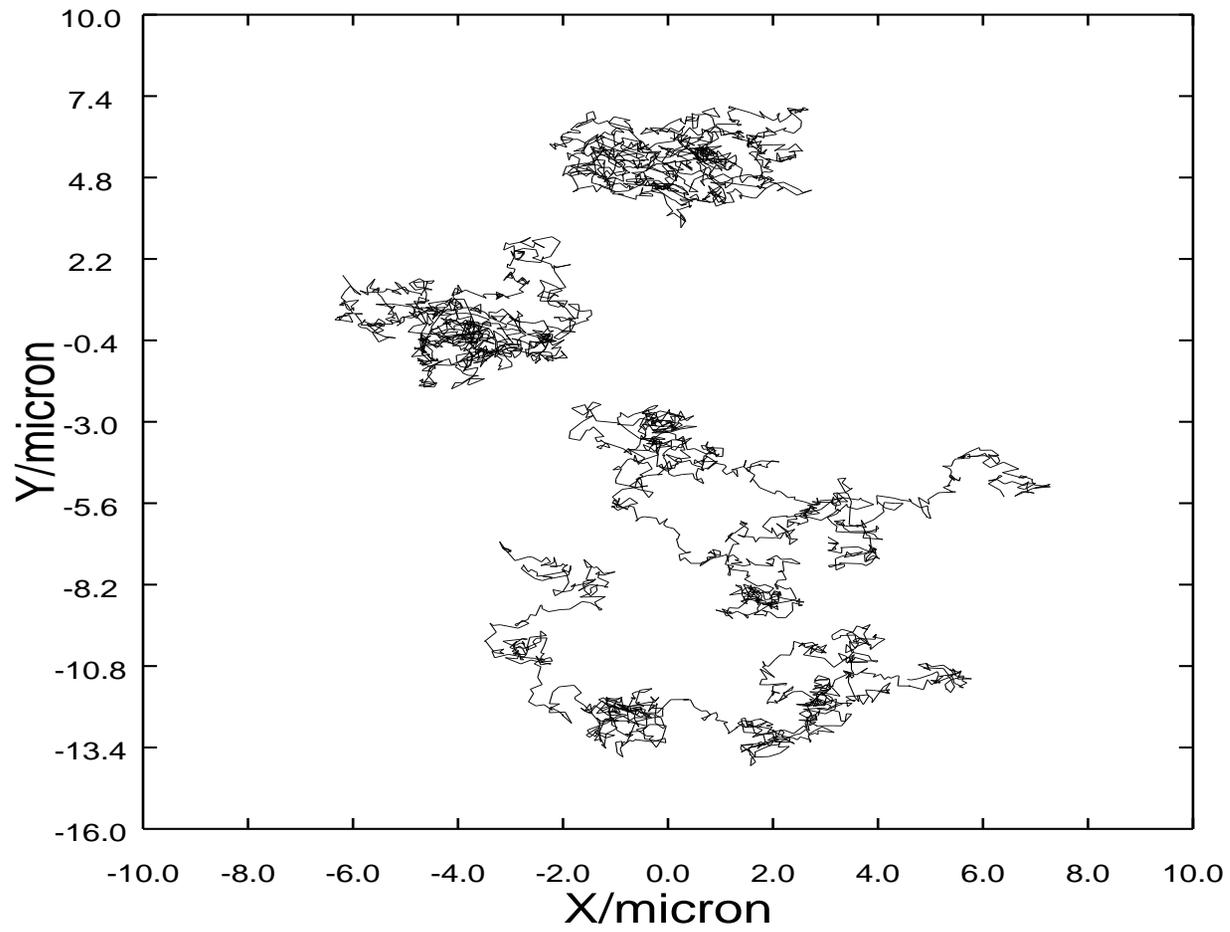
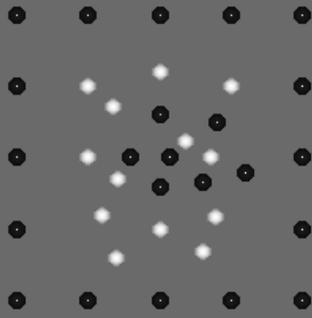
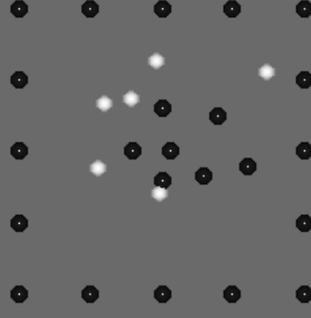
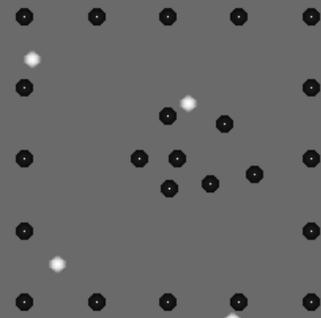
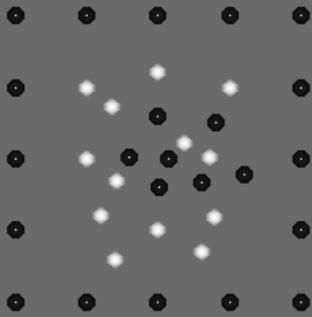
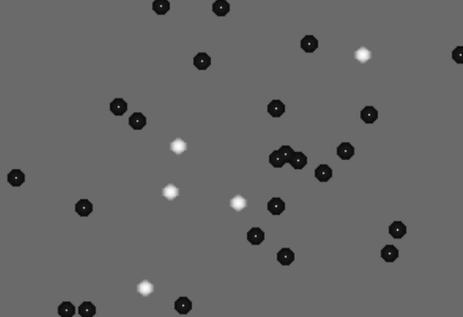
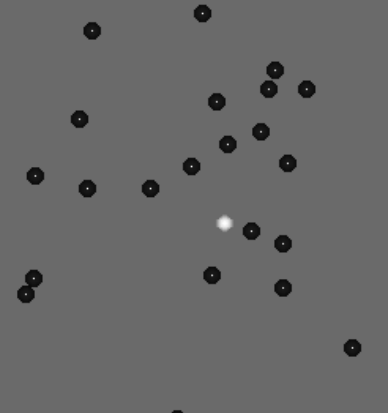
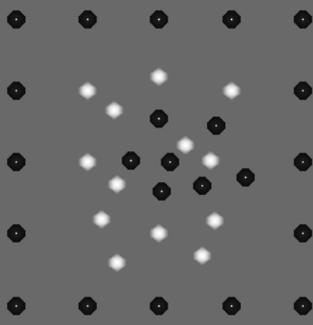
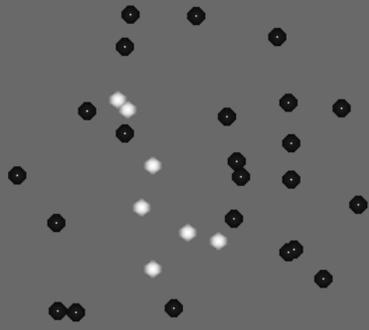
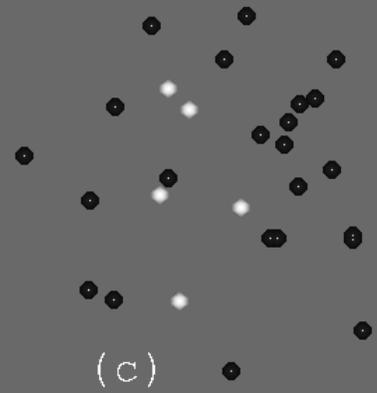
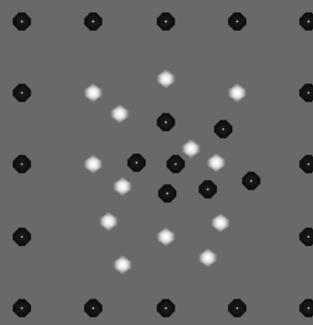
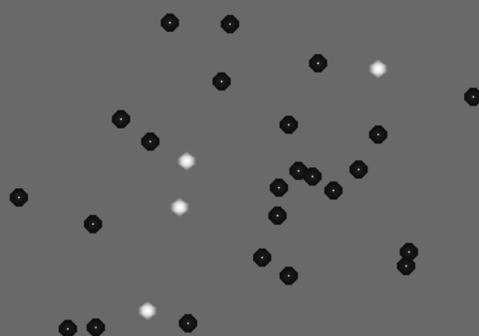
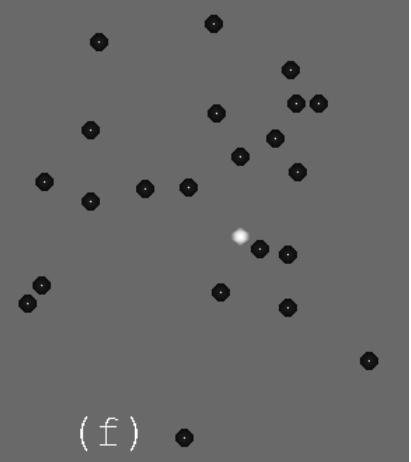
\end{document}